\documentclass[5p,twocolumn]{elsarticle}
\usepackage{graphicx,latexsym}
\usepackage{dcolumn}
\usepackage{amssymb,amsmath,bm}
\usepackage{subfigure}
\usepackage{braket}
\usepackage{siunitx}
\usepackage{array}
\usepackage{physics}
\usepackage{float}
\usepackage[nopar]{lipsum}
\usepackage{caption}
\usepackage{multirow}
\usepackage{bigstrut}

\biboptions{sort&compress}

\usepackage[super]{nth}

\usepackage{hyperref}
\hypersetup{
    pdfnewwindow=true,       % links in new window
    colorlinks=true,         % false: boxed links; true: colored links
    linkcolor=blue,          % color of internal links
    citecolor=blue,          % color of links to bibliography
    filecolor=magenta,       % color of file links
    urlcolor=black           % color of external links
}

\usepackage[normalem]{ulem}

\def\sec#1{Sec.\ \ref{#1}}
\def\eq#1{Eq.\ (\ref{#1})}
\def\fig#1{Fig.\ \ref{#1}}

\journal{}

\begin{document}

\begin{frontmatter}

%-----------------------------------------------------------------

\title{Thermal transport controlled by intra- and inter-dot Coulomb interactions in sequential and cotunneling serially-coupled double quantum dots}
	
\author[a1]{Bashdar Rahman Pirot}
\address[a1]{Physics Department, College of Education, 
	University of Sulaimani, Sulaimani 46001, Kurdistan Region, Iraq}

\author[a2,a3]{Nzar Rauf Abdullah\corref{correspondingauthor}}
\ead{nzar.r.abdullah@gmail.com}
\address[a2]{Division of Computational Nanoscience, Physics Department, College of Science, 
	University of Sulaimani, Sulaimani 46001, Iraq}
\address[a3]{Computer Engineering Department, College of Engineering, Komar University of Science and Technology, Sulaimani, Iraq}

\author[a4]{Andrei Manolescu}
\address[a4]{Reykjavik University, School of Science and Engineering, Menntavegur 1, IS-101 Reykjavik, Iceland}

\author[a5]{Vidar Gudmundsson}
\ead{vidar@hi.is}
\address[a5]{Science Institute, University of Iceland, Dunhaga 3, IS-107 Reykjavik, Iceland}

%----------------------------------------------------------------

\begin{abstract}
	
We study thermoelectric transport through a serial double quantum dot (DQD) coupled to two metallic leads with different thermal energies. We take into account the electron sequential and cotunneling effects via different master equation approaches. In the absence of intra- and inter-dot Coulomb interactions, a small peak in thermoelectric and heat currents is found for $E_{\rm L} \text{=} E_{\rm R}$ indicating the Coulomb blockade DQD regime, where $E_{\rm L}(E_{\rm R})$ is the energy of the state of the left(right) quantum dot. In the presence of intra- and inter-dot Coulomb interactions with strengths U$_{\rm intra}$, and U$_{\rm inter}$, respectively, avoided crossings or resonance energies between the intra- and the inter-dot two-electron states, 2ES, are found. These resonances induce extra transport channels through the DQD leading to strong side peaks in the thermoelectric and heat currents at $ E_{\rm L} \text{-} E_{\rm R} = \pm (U_{\rm intra} \text{-} U_{\rm inter})$ in addition to the main peak generated at $E_{\rm L} \text{=} E_{\rm R}$. The current side peaks are enhanced by increased strength of the Coulomb interactions. 
Interestingly, the current side peaks are enhanced when cotunneling terms are considered in which the resonances of the 2ESs assist the electron cotunneling process through the system. Furthermore, the issue of coherences is carefully checked in the DQD-leads system via different approaches to the master equation, which are the Pauli, the Redfield, a first order Lindblad, and the first- and second order von-Neumann methods. We realize that the Pauli method gives a wrong results for the thermoelectric transport when the role of the coherences is relevant.
\end{abstract}

\begin{keyword}
Serial quantum dots \sep Coulomb interaction \sep Thermal transport \sep Cotunneling \sep Quantum master equations 
\end{keyword}

\end{frontmatter}

\section{Introduction} 

Double quantum dot systems are important and useful for understanding fundamental quantum physics and nanoelectronic-based applications \cite{doi:10.1021/acsanm.0c01386, Bennett2000, RevModPhys.75.1, Wiel_2006}. This is due to it's energy quantization property which makes the DQD to be sensitive to the environment, including phonons, photons, and electron reservoirs \cite{Tagani.1.4767376, GUDMUNDSSON20181672, doi:10.1002/andp.201600177}. As a result, several interesting phenomena such as phonon- \cite{doi:10.1021/acs.nanolett.0c04017}, and photon-assisted electrical \cite{vanderWiel2002, PhysicaE.64.254} and thermal \cite{doi:10.1021/acsphotonics.5b00532} transport have been studied in quantum dots. 
The phonon-assisted transport enables the conversion of local heat into electrical power in nanosized heat engines, and the photon-assisted transport induces extra current peaks via formation of extra energy levels within the Coulomb blockade quantum dot coupled to a photon source \cite{Kouwenhoven50.2019, Wharam1995}. This thus demonstrates the feasibility to use quantum dots as sensitive and frequency selective detectors in the microwave frequency range. 

The Coulomb interactions can also assist the transport of electrons through quantum dot systems, inducing interesting phenomena such as thermoelectric current and Coulomb-blockade plateaus \cite{TORFASON2013178}, and Coulomb drag effects \cite{Jong2018}. The drag effect is a phenomenon, where a current flowing in a so-called drive conductor induces a voltage across a nearby drag conductor via the Coulomb interaction \cite{RevModPhys.88.025003}. It has been shown by experimental and theoretical arguments that the 
cotunneling process is essential to obtain a correct qualitative understanding of the transport and drag properties in a Coulomb-coupled double quantum dot \cite{PhysRevLett.117.066602, Talbo2017}. In addition, a dynamical signature of the Coulomb-blockade is seen, where the excited states are more active than the ground state even at a low electron number \cite{PhysRevB.81.155442}.

The intra- and inter-types of Coulomb interactions can together or individually have significant effects on the electron transport through quantum dots \cite{WU2011749}.
In the presence of intra-Coulomb interaction in a quantum dot connected to another dot,  a nonvanishing current through an unbiased Coulomb blocked dot is found due to 
the momentum transfer between the two dot systems. Intradot transitions thus appear
when the level spacing coincides with the transferred energy. In this case a strong drag effect is seen \cite{Moldoveanu_2009}. On the other hand, in the presence of both types of Coulomb interactions and the influence of different parts of the Coulomb interaction inside a triple quantum dots have been identified. It has been demonstrated that the Coulomb interaction between electrons opens up a large variety of different channels, when the spin degeneracy of the levels is included. In particular, it enhances electron transport in the triple dots \cite{Goldozian2016}.

Motivated by the aforementioned studies, we consider double serial quantum dots coupled to two metallic leads via tunneling contacts. The inter- and intra-Coulomb interaction in the DQD are taken into account, and the Coulomb interaction in the leads is ignored. We find that the most relevant states in the transport are two-electron states, 2ES. Two different categories of 2ES are seen. 
First, the 2ESs within a dot and influenced by the intra-dot Coulomb interaction. Second, the 2ESs formed between dots and influenced by the inter-dot Coulomb interaction. A pronounce enhancement in thermoelectric transport 
is found when these two types of 2ESs interact. In addition, different types of master equations are used to study the thermoelectric, the heat, and the energy currents, and their results are compared. 
Finally, the role of cotunneling via a second-order von-Nummen equation is shown on the thermoelectric current.

The work is organized as follows: in \sec{section_model} the Hamiltonian of the total system, QD and leads, and the master equation formalism are presented. In \sec{section_results} the results of thermal transport under Coulomb interaction are displayed. In \sec{section_conclusion}, the main conclusion and remarks are presented.

\section{Model and Theory}\label{section_model}
The system under study is a composed of a serial DQD connected to two external metallic leads via tunnel barriers which is described by the following Hamiltonian \cite{Goldozian2016, Abdullah2017}
\begin{equation}
\hat{H} = \hat{H}_{\rm DQD} + \hat{H}_{\rm leads} + \hat{H}_{\rm T},
\end{equation}
where $\hat{H}_{\rm DQD}$ is the Hamiltonian of the DQD, $\hat{H}_{\rm leads}$ indicates the Hamiltonian of the leads, and $\hat{H}_{\rm T}$ represents the tunneling Hamiltonian between the DQD and the leads. The serial DQD is connected to the two leads from the left and the right sides.
The Hamiltonian of DQD can be described via \cite{doi:10.1021/acs.nanolett.0c04017}.
\begin{equation}
	\hat{H}_{\rm DQD} = \hat{H}_{\rm L} + \hat{H}_{\rm R} +  \hat{H}_{\rm \Omega},
\end{equation}
where $\hat{H}_{\rm L}$ is the Hamiltonian of the left quantum dot, $L$, $\hat{H}_{\rm R}$ is for the right dot, $R$, and $ \hat{H}_{\rm \Omega}$ is the coupling between the dots.
The Hamiltonian of the left dot in occupation number representation can be defined as
\begin{align}
\hat{H}_{\rm L} &= \sum_{n\sigma} E_{\rm L} \, \hat{a}^{\dagger}_{Ln\sigma} \hat{a}_{Ln\sigma}
                \nonumber \\
	            & + \frac{U_{\rm intra}}{2} \sum_{n n^{\prime} m m^{\prime}} \sum_{\sigma \sigma^{\prime}}
	            \hat{a}^{\dagger}_{Ln\sigma} \hat{a}^{\dagger}_{Ln^{\prime}\sigma^{\prime}}
	            \hat{a}_{Lm^{\prime}\sigma^{\prime}} \hat{a}_{Lm\sigma}.
\end{align}
Herein, $E_{\rm L}$ is the energy level of the left dot, $\hat{a}^{\dagger}_{Ln\sigma}$($\hat{a}_{Ln\sigma}$) denotes the spin-dependent, $\sigma$, creation(annihilation) operator in the left dot, and $U_{\rm intra}$ is a constant intradot Coulomb interaction. The right dot is treated analogously with $L \rightarrow R$. The two quantum dots are coupled by
\begin{align}
\hat{H}_{\Omega} & = \sum_{nn^{\prime}, \sigma} \Omega \, \hat{a}^{\dagger}_{Rn^{\prime}\sigma} \hat{a}_{Ln\sigma} + h.c. \nonumber \\
                 & + \sum_{nn^{\prime}, \sigma\sigma^{\prime}} U_{\rm inter} \,
                  \hat{a}^{\dagger}_{Ln{\sigma}}  \hat{a}^{\dagger}_{Rn^{\prime}{\sigma^{\prime}}}
                  \hat{a}_{Rn^{\prime}{\sigma^{\prime}}}  \hat{a}_{Ln{\sigma}},
\end{align}
with the inter-dot tunnel coupling strength, $\Omega$, and the inter-dot Coulomb interaction, $U_{\rm inter}$, between neighboring dots. Both the intra- and the inter-Coulomb matrix elements are treated in the same way as is shown in Ref.\ \cite{Goldozian2016}.
The Hamiltonian of each lead is
\begin{equation}
\hat{H}_{\rm leads} = \sum_{q \sigma l} \varepsilon_{q \sigma l} \, d^{\dagger}_{q \sigma l} \, d_{q \sigma l}, 
\end{equation}
where $\varepsilon_{q \sigma l}$ denotes the spin-dependent energy levels of the leads, $d^{\dagger}_{q \sigma l}$($d_{q \sigma l}$) indicates the electron creation(annihilation) operator in the leads with index $l$ (typically right and left lead, respectively), and the momentum quantum number $q$ refers to states in a continuum of states.
The tunneling Hamiltonian, $\hat{H}_{\rm T}$, can be expressed by 
\begin{equation}
\hat{H}_{\rm T} = \sum_{n,q \sigma l} t_{nl} \, a^{\dagger}_{n\sigma} d_{q \sigma l} + {\rm H.c.},
\end{equation}
where $t_{nl}$ introduces the tunneling amplitude between the leads and the $L$ or the $R$ dot. 
The tunneling amplitude is connected to the tunneling rate via \cite{KIRSANSKAS2017317}
\begin{equation}
	\Gamma_{\alpha q, i}(E) = 2\pi \sum_{q} |t_{n l}|^2 \delta(E_{L,R}-\varepsilon_{q \sigma l}).
	\label{gamma_1}
\end{equation}

The electron evolution in the total system, the DQD and the leads, is described by the
Liouville-von Neumann equation for the total density matrix of the full system, $\rho(t)$ \cite{Haake1973,Breuer2002}. Since we are interested in the time evolution of the electrons 
in the DQD, the dynamics of the total system is projected on the central system, the serial
DQD, leading to the reduced density operator, $\rho_{\rm S}(t)$, describing the electrons in the DQD under the influence of the leads. This can be obtained by tracing out the variables of the leads \cite{Nzar_2016_JPCM, ABDULLAH2018}
\begin{equation}\label{master_equation}
	\rho_\mathrm{S}(t) = Tr_\mathrm{leads} \left\{ \rho (t) \right\}.
\end{equation}
The solution of the resulting \eq{master_equation} for the reduced density operator is implemented with the QmeQ package \cite{KIRSANSKAS2017317}, which 
numerically gives $\rho_\mathrm{S}(t)$ in the interacting many-body Fock basis of the DQD. 
Consequently, the transport properties of the system can be obtained using $\rho_\mathrm{S}(t)$.

We calculate the thermoelectric properties of the DQD in the steady state regime. So, we apply a thermal gradient to the DQD via a temperature difference between the leads \cite{PhysRevB.90.115313-2}.
The thermoelectric current, TEC (I$^{\rm TEC}$), through the lead channel $n$ is introduced as
\begin{equation}
	I^{\rm TEC}_{n} =  -\frac{\partial}{\partial t} \expval{N_{n}},
	= -i \expval{[H, N_{n}]},
\end{equation}
with $N_{n} = \sum_q d^{\dagger}_{n q} d_{n q}$. 
The energy current, EC ($I^{\rm EC}_{n}$), is determined using
\begin{equation}
	I^{\rm EC}_{n}  = -\frac{\partial}{\partial t} \expval{H_n}
	= -i\expval{[H, H_{n}]},
\end{equation}
where $H_{n} = \sum_q d^{\dagger}_{n q} d_{n q}$.
Finally, the heat current, HC ($I^{\rm HC}_{n}$), \cite{ABDULLAH2018199} emerging from the lead channel $n$ 
is represented by 
\begin{equation}
	I^{\rm HC}_{n} = I^{\rm EC}_{n} - \mu \, I^{\rm TEC}_{n},
	\label{HC_EC}
\end{equation}
where $\mu$ is the common chemical potential of the leads.

\section{Results}\label{section_results}
In this section, we introduce the results for a device, a serial DQD with small interdot tunnel coupling $\Omega = 0.05$~meV, that is weakly coupled to the leads. The temperature gradient applied to the leads generates different energy distributions in the leads around their electrochemical potentials $\mu$. The chemical potential of both leads are assumed to be equal, $\mu_L = \mu_R = 0.0$.
The the first-order von-Neumann master equation, 1vN, implemented with the QmeQ package is used to calculate the thermoelectric properties of the DQD-leads system with the assumption 
that $k_B = 1.0$. The evolution of the central system, the serial DQD, is defined by the reduced density operator, $\rho_\mathrm{S}$ \cite{GUDMUNDSSON20181672, 0953-8984-30-14-145303},
where the density matrix elements for the sectors in the Fock space determined by the number of electron/hole excitation in the leads are grouped together \cite{Goldozian2016}.
In the 1vN method, only sequential tunneling in the case of coherences is included, 
and the coupling strength between the DQD and leads has to be smaller than the temperature of the leads $\Gamma_{L,R} << T_{L,R}$. Another remark for the 1vN is that it can violate the positivity of $\rho_\mathrm{S}$ if the coupling of the DQD with leads is too strong, which is an unphysical situation that must be avoided  \cite{Goldhaber-Gordon1998}. 

In order to satisfy the conditions for 1vN, we assume that the thermal energy of the leads $k_BT_{\rm L}(k_BT_{\rm R})$ is $1.5$($0.5$)~meV, and the coupling strength between the DQD and the leads is $\Gamma_{\rm L,R} = 90\times 10^{-6}$~meV. The condition of $\Gamma_{L,R} << T_{L,R}$ is thus satisfied in our calculations \cite{Kouwenhoven_2001}.

\subsection{The DQD without Coulomb interactions} 
\begin{figure}[htb]
	\centering
	\includegraphics[width=0.4\textwidth]{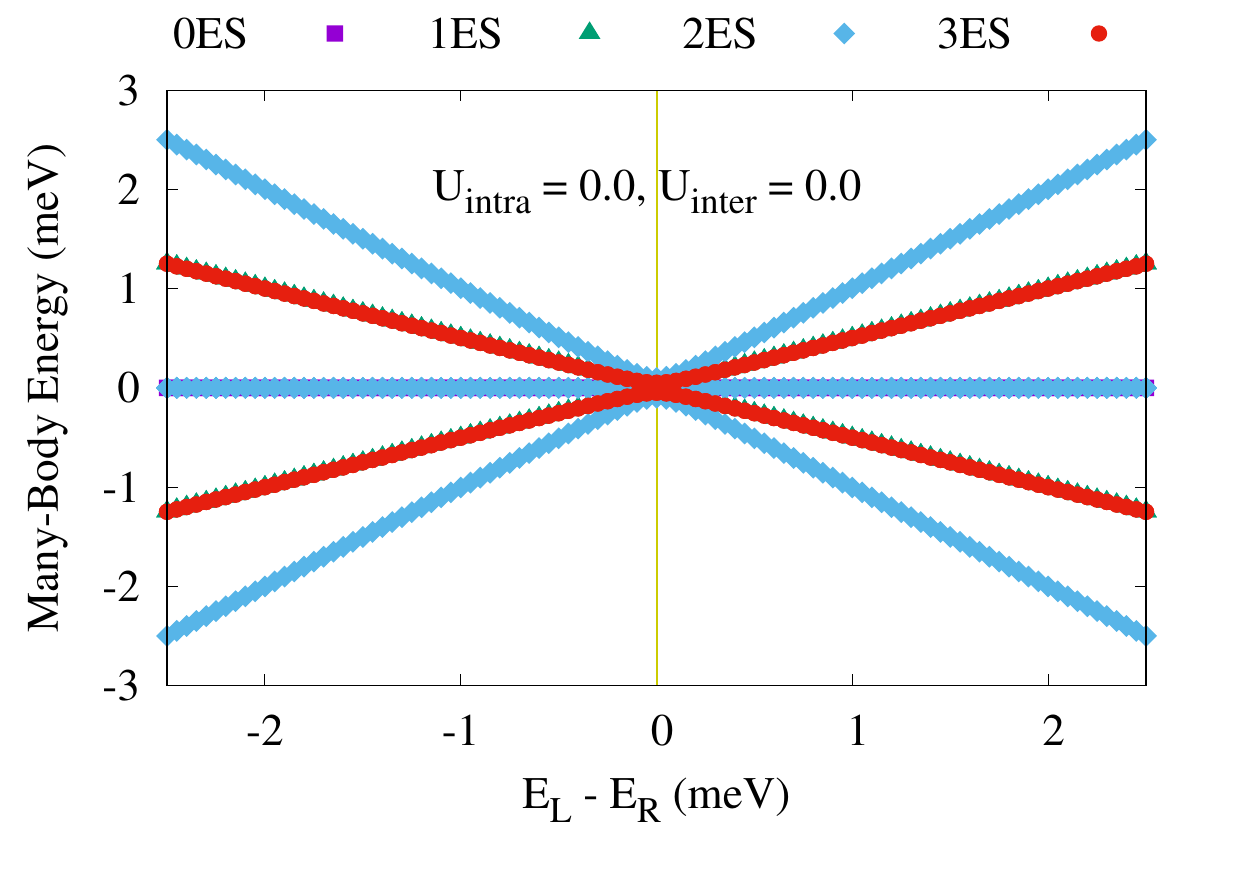}  %MBE_Uintra_0_Uinter_0
	\caption{Many-Body (MB) energy spectrum as a function of $E_{\rm L}\text{-}E_{\rm R}$ for zero-electron states, 0ES, (purple rectangles),  one-electron state, 1ES, (green triangle), two-electron states, 2ES (blue diamond), and three-electron states, 3ES, (red circles) for the DQD without Coulomb interactions, U$_{\rm intra}$ = U$_{\rm inter} = 0.0$. The vertical golden line is the position when $E_{\rm L} = E_{\rm R}$.
	There are two types of 2ES: They can have both electrons
	in the same dot (two lines with blue diamonds that change with $E_{\rm L}\text{-}E_{\rm R}$) or one electron in either dot (the blue diamonds that do not change with $E_{\rm L}\text{-}E_{\rm R}$).}
	\label{fig01}
\end{figure}

We first consider no Coulomb interactions, intra- and inter, in the DQD, i.e.\ $U_{\rm intra} = U_{\rm inter} = 0.0$. The energy spectrum of the DQD in the absence of the Coulomb interactions is shown in \fig{fig01}. The Many-body energy spectrum is displayed as a function of $\Delta = E_{\rm L}\text{-}E_{\rm R}$. One can clearly see that the zero-electron state, 0ES, (purple rectangles) remains at zero independent of the tuning of $\Delta$. In contrast, the one-electron states, 1ES, (green rectangles), and the three-electron states, 3ES, (red circles) coincide and change with $\Delta$. Each 1ES, and 3ES is double degenerate due to the spin of the electrons, $\sigma = \, \uparrow, \downarrow$. There are two types of 2ES, states with one electron in each dot have a flat dispersion in Fig.\ \ref{fig01}, and the states with both electrons in the same dot
have a strong dispersion.
The vertical golden line denotes the position where the $E_{\rm L}$ is equal to $E_{\rm R}$. At this point the energy levels of the left dot are crossing the energy levels of the right dot. It is interesting to see that the 1ES, 2ES, and 3ES of the left dot are in resonance with those of the right dot at $E_{\rm L} = E_{\rm R}$.

\begin{figure}[htb]
	\centering
	\includegraphics[width=0.45\textwidth]{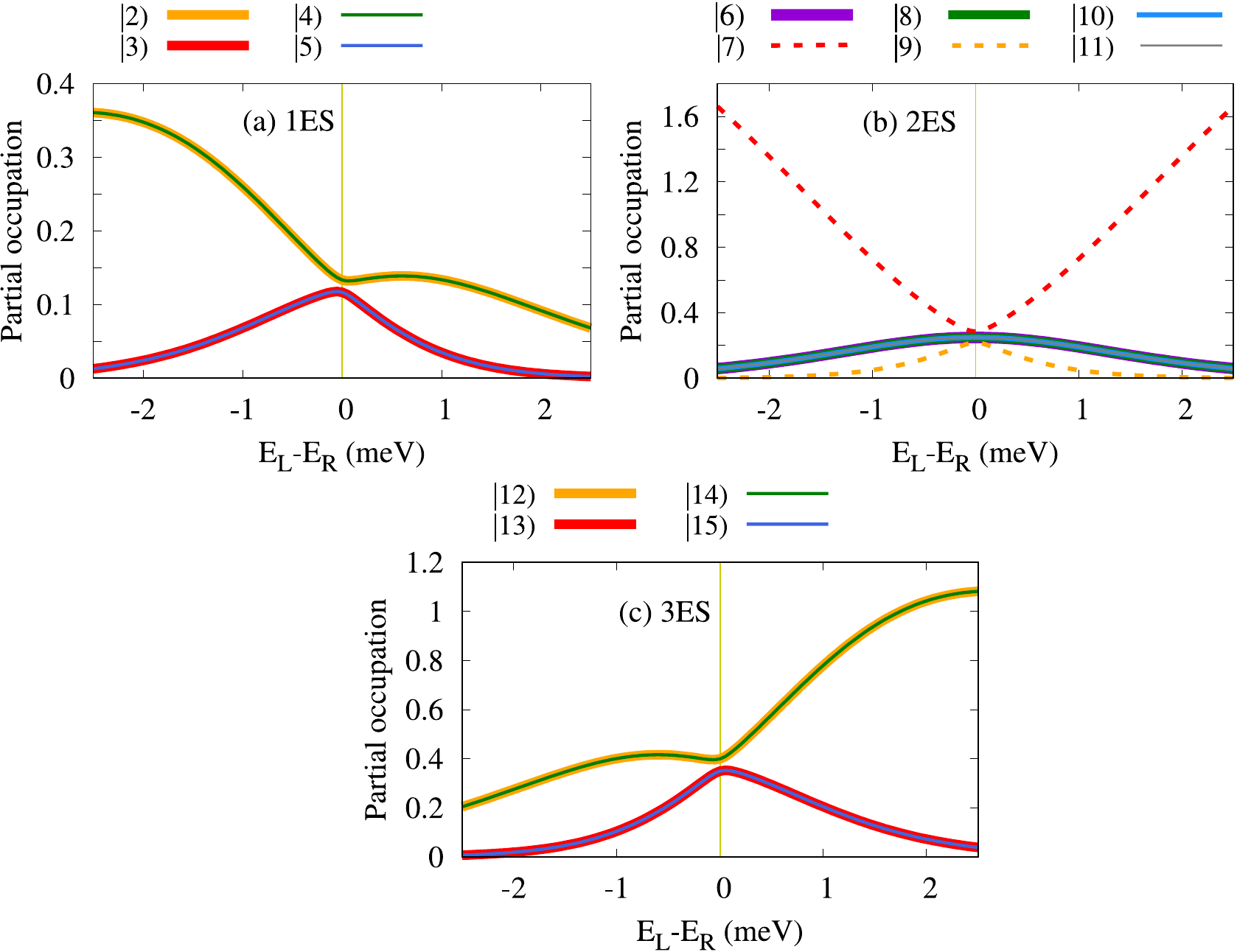}
	\caption{Partial occupation as a function of $E_{\rm L}\text{-}E_{\rm R}$ for the one-electron state, 1ES, (a), the two-electron states, 2ES (b), and the three-electron states, 3ES, (c) of the DQD without Coulomb interactions, $U_{\rm intra} = U_{\rm inter} = 0.0$. The chemical potentials of the leads are $\mu_{\rm L} = \mu_{\rm R} = 0.0$, which coincides with the vertical golden line when $E_{\rm L}=E_{\rm R}$.
	In Fig. (b), dashed lines are the occupation of the 2ES when both electrons are found in the same dot, and solid lines are the occupation of the 2ES when one electron is in either dot. The thermal energy of the leads are assumed to be $k_BT_L = 1.5$~meV and $k_BT_R = 0.5$~meV, and 
	the coupling strength is $\Gamma_{L,R} = 90\times10^{-6}$~meV. }
	\label{fig02}
\end{figure}

To get insight into the physical properties, we present the occupation or the partial occupation of the energy states of the DQD in \fig{fig02} for the 1ESs (a), 2ESs (b), and 3ESs (c). 
We find four occupied 1ESs shown in \fig{fig02}(a), which are labeled as 
$|2)$, $|3)$, $|4)$, and $|5)$. 
The state $|1)$ is the 0ES which has zero occupation (not shown). For negative values of $E_{\rm L}\text{-}E_{\rm R}$, which means that $E_{\rm L}$ is lower than $E_{\rm R}$, the states $|2)$ and $|4)$ are occupied indicating an electron shift from left lead to the left dot. These two states thus correspond to the left dot. 
At the same time, occupation of $|3)$ and $|5)$ is almost zero indicating unoccupied states of the right quantum dot. Further tuning the $E_{\rm L}\text{-}E_{\rm R}$, the states of $|2)$ and $|4)$ are slowly loosing their occupation while the states of $|3)$ and $|5)$ gain occupation until $E_{\rm L}=E_{\rm R}$ where the occupation of these four states are almost the same demonstrating a resonance condition of these states. This confirms the crossing or resonance condition of the 1ESs shown in \fig{fig01} at $E_{\rm L}=E_{\rm R}$. Further tuning $E_{\rm L}\text{-}E_{\rm R}$ to positive values, meaning that $E_{\rm L}$ is higher than $E_{\rm R}$, all the 1ESs loose occupation or become unpopulated.

The same scenario can be seen for the 3ES shown in \fig{fig02}(c). The difference here is that the occupation of the 3ESs at negative value of $E_{\rm L}\text{-}E_{\rm R}$ is lower than that for positive value of $E_{\rm L}\text{-}E_{\rm R}$.

The most interesting part is the occupation of 2ESs presented in \fig{fig02}(b) where the dashed lines are the occupation of the 2ESs with both electrons in the same dot, while the solid lines 
are the occupation of the 2ES when one electron is in either dot. We can clearly see that 
both types of 2ESs are in resonance at $E_{\rm L}=E_{\rm R}$. At this point, the 2ESs corresponding to one electron in either dot have gained occupation.

\begin{figure}[htb]
	\centering
	\includegraphics[width=0.35\textwidth]{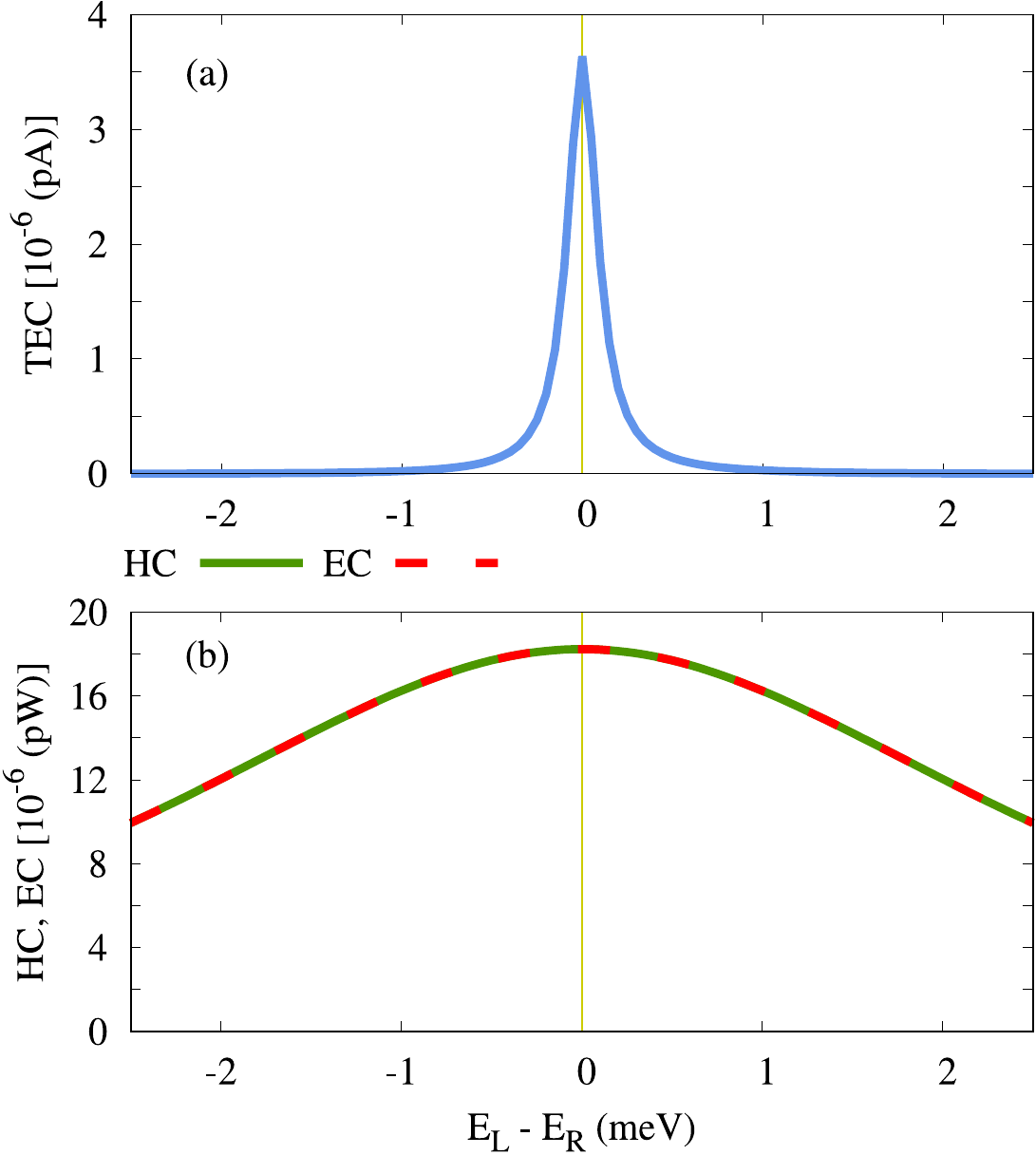}
	\caption{Thermoelectric current, TEC, Heat current, HC, Energy current, EC, as a function of $E_{\rm L}\text{-}E_{\rm R}$ for the double quantum dots without intra- and inter-dot Coulomb interactions, $U_{\rm intra} = 0.0$ and $U_{\rm inter} = 0.0$, respectively. The chemical potentials of the leads are $\mu_L = \mu_R = 0.0$~meV which coincides with the vertical golden line when $E_L = E_R$. The thermal energy of the leads are assumed to be $k_BT_L = 1.5$~meV and $k_BT_R = 0.5$~meV, and the coupling strength is $\Gamma_{L,R} = 90\times10^{-6}$~meV. }
	\label{fig03}
\end{figure}

The occupation of all states generates or controls the themoelectric current (a), the heat and energy currents (b) shown in \fig{fig03}, and the currents get maxima at $E_{\rm L}=E_{\rm R}$ corresponding to the crossing of energy states of the DQD.
The thermal currents are very small with the TEC in the range of atto amperes and the HC, and EC are in the range of atto watts. This could occur as the chemical potentials of the leads are set to zero. At this point, the states of the DQD are aligned with the chemical potential at $E_L\text{-}E_R = 0.0$~meV. This location thus corresponds to half filling of states, and at the half filling the thermal currents should be very small or zero \cite{TAGANI201336_PAT}. The HC is exactly equal to the EC in the DQD because the chemical potential is zero. At zero value of $\mu$, $I^{\rm HC}$ should be equal to $I^{\rm EC}$ as can be predicted from \eq{HC_EC}.

\subsection{The DQD with Coulomb interactions} 

In this section, we consider the intra- and inter-dot Coulomb interaction in the DQD. The strength of the inter-dot Coulomb interaction is fixed as U$_{\rm inter} = 1.0$~meV, and three values of the intra-dot Coulomb interaction will be considered, U$_{\rm intra} = 1.8$, $2.2$, and $2.6$~meV. Under these assumptions, U$_{\rm inter}$, U$_{\rm intra} > k_B T$ indicating that the charging energy is larger than the thermal energy of the leads. So, the Coulomb interactions can effectively influence the characteristics of the thermal transport \cite{PhysRevB.53.12625}. 

\begin{figure}
	\centering
	\includegraphics[width=0.35\textwidth]{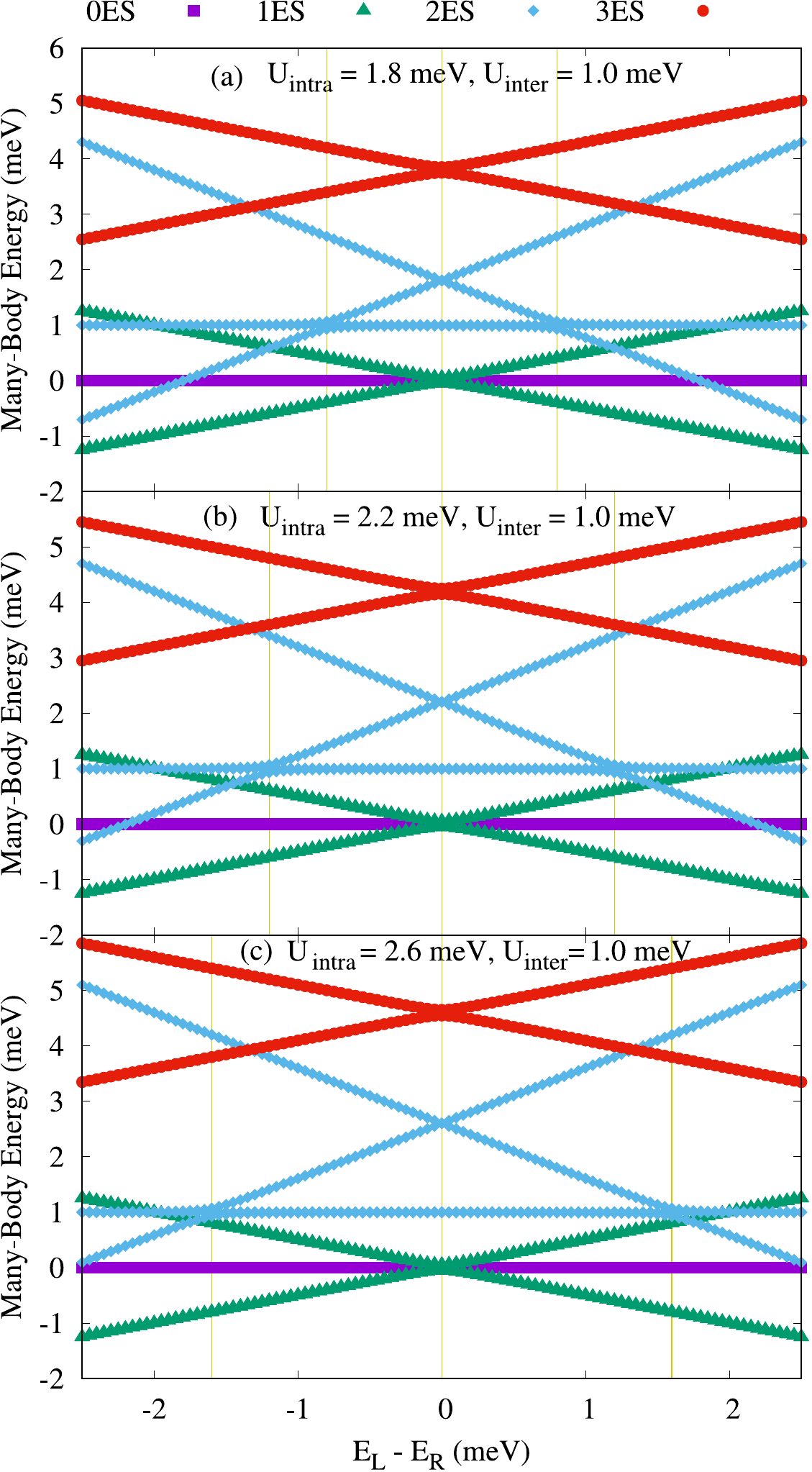}  %MBE_Uintra_0_Uinter_0
	\caption{Many-Body (MB) energy spectrum as a function of $E_{\rm L}\text{-}E_{\rm R}$ for zero-electron states, 0ES, (purple rectangles), one-electron states, 1ES, (green triangle), two-electron states, 2ES (blue diamond), and three-electron states, 3ES, (red circles). The vertical golden line (middle line) is the energy position when $E_L = E_R$, and the other two vertical golden lines (left and right) are the positions when the two types of the 2ES are crossing. Three different strengths of intra- and inter-dot Coulom interactions are considered. The strength of the inter-dot Coulomb interaction is fixed at U$_{\rm inter} = 1.0$~meV, and the strength of intra-dot Coulomb interaction is changed to U$_{\rm intra} = 1.8$ (a), $2.2$ (b), $2.6$~meV (c). The thermal energy of the leads are assumed to be $k_BT_L = 1.5$~meV and $k_BT_R = 0.5$~meV, and the coupling strength is $\Gamma_{L,R} = 90\times10^{-6}$~meV. }
	\label{fig04}
\end{figure}

The many-body energy (MBE) spectrum of the serial DQD including intra-dot Coulomb interactions with strength of U$_{\rm intra} = 1.8$ (a), $2.2$ (b), $2.6$~meV (c) is presented in \fig{fig04}. The inter-dot Coulomb interaction strength is assumed to be U$_{\rm inter} = 1.0$~meV here. Compairing to the energy spectrum of the DQD without Coulomb interactions shown in \fig{fig01}, several channels due to crossing of energy states emerge in the MBE spectrum, when the Coulomb interactions are considered. 

The first channel is the position when $E_{\rm L}\text{-}E_{\rm R} = 0$. At this position, three crossings of the states 1ESs, 2ESs and 3ESs of both quantum dots are formed at different energy values of the MBE spectrum. 
The crossing of the 1ESs at the zero value of the MBE is seen, while the crossings of intra-dot 2ESs appears at $1.8$, $2.2$, and $2.6$~meV of the MBE for U$_{\rm intra} = 1.8$, $2.2$, and $2.6$~meV, respectively.
Furthermore, the inter-dot 2ESs emerges at $1.0$~meV for all three cases of the intra-dot Coulomb interaction, which is expected as U$_{\rm inter} = 1.0$~meV. The crossings of the 3ESs are found at $3.8$, $4.2$, and $4.6$~meV of the MBE for U$_{\rm intra} = 1.8$, $2.2$, and $2.6$~meV at $E_{\rm L}\text{-}E_{\rm R}=0$, respectively. 
These three crossing of the 1ESs, 2ESs, and 3ESs at $E_{\rm L}\text{-}E_{\rm R}=0$ are identified as the main channel for electron transport.

The second and the third channels are seen at the crossing energy between inter- and intera-dot 2ESs at $E_{\rm L}\text{-}E_{\rm R}= \pm \, (U_{\rm intra} \text{-} \, U_{\rm inter})$. As a results, extra side-channels are found at $\pm 0.8$, $\pm 1.2$, and $\pm 1.6$~meV for U$_{\rm intra} = 1.8$, $2.2$, and $2.6$~meV, respectively. These extra-side channels are totally formed by the presence of the intra- and the inter-dot Coulomb interactions in the DQD. These extra-side channels are not seen in for the transport results in the absence of the Coulomb interactions shown in \fig{fig01}.

\begin{figure}
	\centering
	\includegraphics[width=0.45\textwidth]{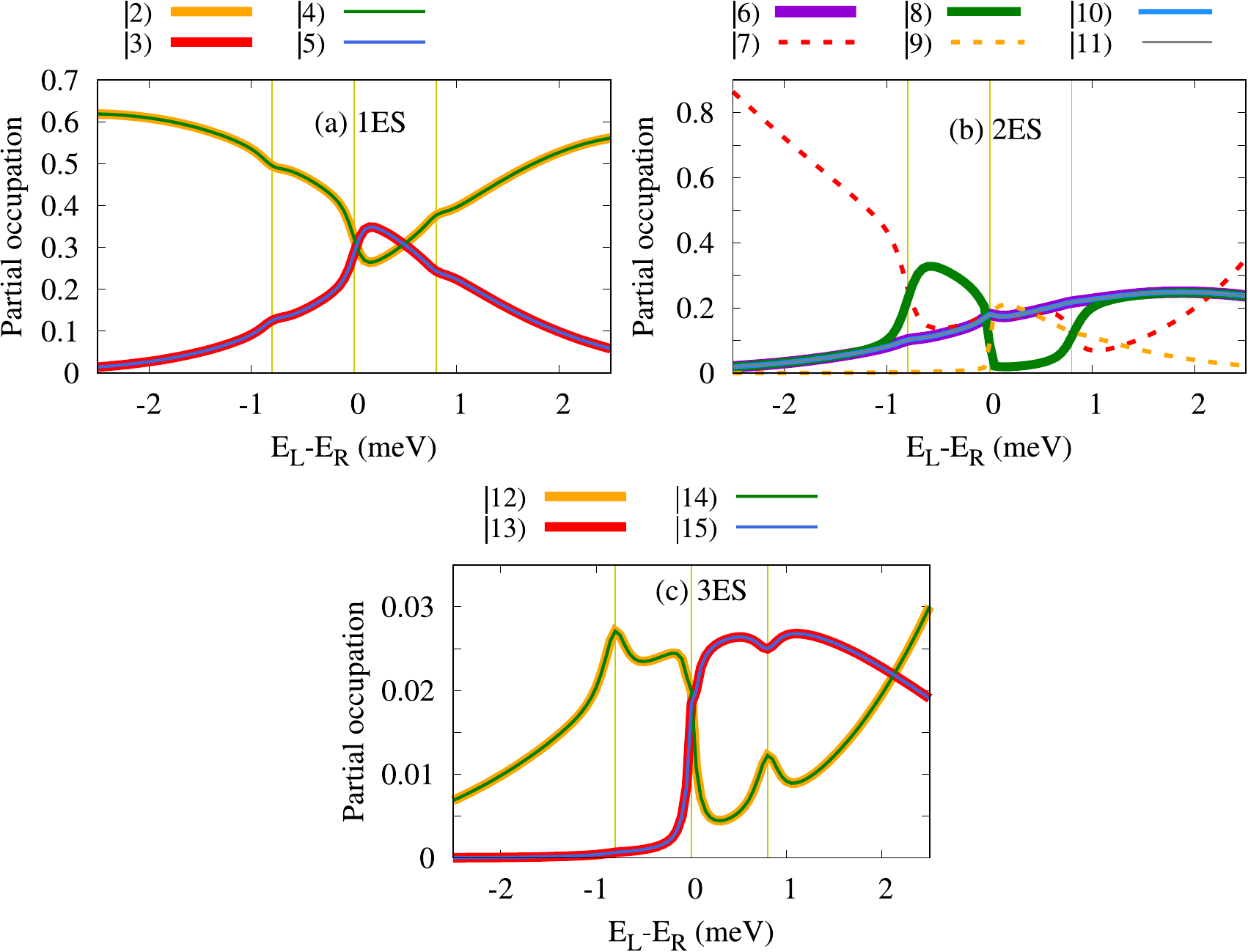}  %MBE_Uintra_0_Uinter_0
	\caption{Partial occupation as a function of $E_L-E_R$ for the one-electron state, 1ES, (a), two-electron states, 2ES (b), and three-electron states, 3ES, (c) in the presence of Coulomb interactions, $U_{\rm intra} = 1.8$~meV and $U_{\rm inter} = 1.0$~meV. The chemical potentials of the leads are $\mu_L = \mu_R = 0.0$~meV, which coincides with the middle vertical golden line when $E_L = E_R$.
	In Fig. (b), dashed lines are the occupation of the 2ES when both electrons are found in the same dot, and solid lines are the occupation of the 2ES when one electron is in either dot. The thermal energy of the leads are assumed to be $k_BT_L = 1.5$~meV and $k_BT_R = 0.5$~meV, and the coupling strength is $\Gamma_{L,R} = 90\times10^{-6}$~meV.}
	\label{fig05}
\end{figure}

To further understand the properties of the serial DQD with Coulomb interactions, their occupation for the 1ESs (a), the 2ESs (b), and the 3ESs (c) are displayed in \fig{fig05} for U$_{\rm inter} = 1.0$~meV and U$_{\rm intra} = 1.8$~meV. The crossings of the 1ESs and the 3ESs are found at $E_{\rm L}\text{-}E_{\rm R} = 0$ in which the occupations of 3ESs are strongly suppressed.

It is very interesting to see the occupations of the intra-dot 2ESs (dashed line) and the inter-dot 2ESs (solid line). Comparing to the occupation of the 2ESs in the absence of Coulomb interaction shown in \fig{fig02}b, the crossing of occupations between the intra-dot 2ESs and the inter-dot 2ESs are found at $E_{\rm L}\text{-}E_{\rm R}= \pm \, (U_{\rm intra} \text{-} \, U_{\rm inter}$). 
At the point where $E_{\rm L}\text{-}E_{\rm R} = \pm\, (U_{\rm intra} \text{-} \, U_{\rm inter}$) = $\pm 0.8$ meV, the populated inter-dot 2ES (green line) and the depopulated intra-dot 2ES (red dashed line) are crossing due to their resonance at $\pm 0.8$ meV shown in \fig{fig04}(a). The ``exchange'' of the occupation of these two types of 2ESs is seen at $E_{\rm L}\text{-}E_{\rm R} = 0.0$.

\begin{figure}
	\centering
	\includegraphics[width=0.35\textwidth]{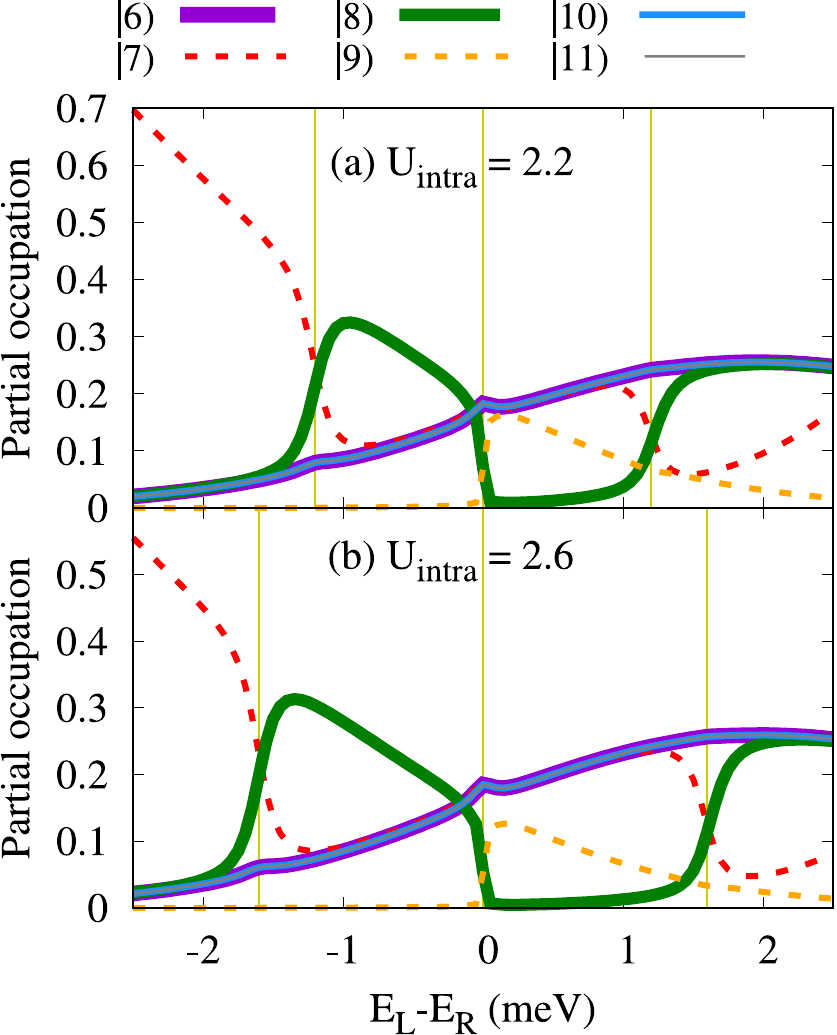}  %MBE_Uintra_0_Uinter_0
	\caption{Partial occupation as a function of $E_L-E_R$ for two-electron states, 2ES where $U_{\rm intra} = 2.2$ (a), and $2.6$~meV (b) with $U_{\rm inter} = 1.0$~meV. The chemical potentials of the leads are $\mu_L = \mu_R = 0.0$~meV which coincides with the middle vertical golden line when $E_L = E_R$.
	The dashed lines are the occupation of 2ES when both electrons are found in the same dot, and solid lines are the occupation of 2ES when one electron in either dot is realized.  The thermal energy of the leads are assumed to be $k_BT_L = 1.5$~meV and $k_BT_R = 0.5$~meV, and the coupling strength is $\Gamma_{L,R} = 90 \times 10^{-6}$~meV.}
	\label{fig06}
\end{figure}

The partial occupation of the the 2ESs in the presence of the intra-dot Coulomb interaction with strength U$_{\rm intra} = 2.2$ (a), and $2.6$~meV (b), and the fixed value of U$_{\rm inter} = 1.0$~meV are displayed in \fig{fig06}. In this case, we only present the occupation of the 2ESs because the occupation of the 1ESs and the 3ESs are qualitatively similar to that shown in \fig{fig05}(a,c).
It is worth to notice that the exchange of occupation between the intra-dot 2ESs and the inter-dot 2ESs are shifted to a higher/lower value of $E_{\rm L}\text{-}E_{\rm R}$ as the strength of intra-dot Coulomb interaction is larger. In the case of U$_{\rm intra} = 2.2$ meV, the exchanges between the depopulated and the populated 2ESs is shifted to $E_{\rm L}\text{-}E_{\rm R} = \pm\, (U_{\rm intra} \text{-} \, U_{\rm inter}$) = $\pm 1.2$ meV, while the exchange takes place at $\pm 1.6$~meV for U$_{\rm intra} = 2.6$ meV. This further confirms the resonance between both types of 2ESs at $\pm\, (U_{\rm intra} \text{-} \, U_{\rm inter}$) for different values of $U_{\rm intra}$.

\begin{figure}
	\centering
	\includegraphics[width=0.45\textwidth]{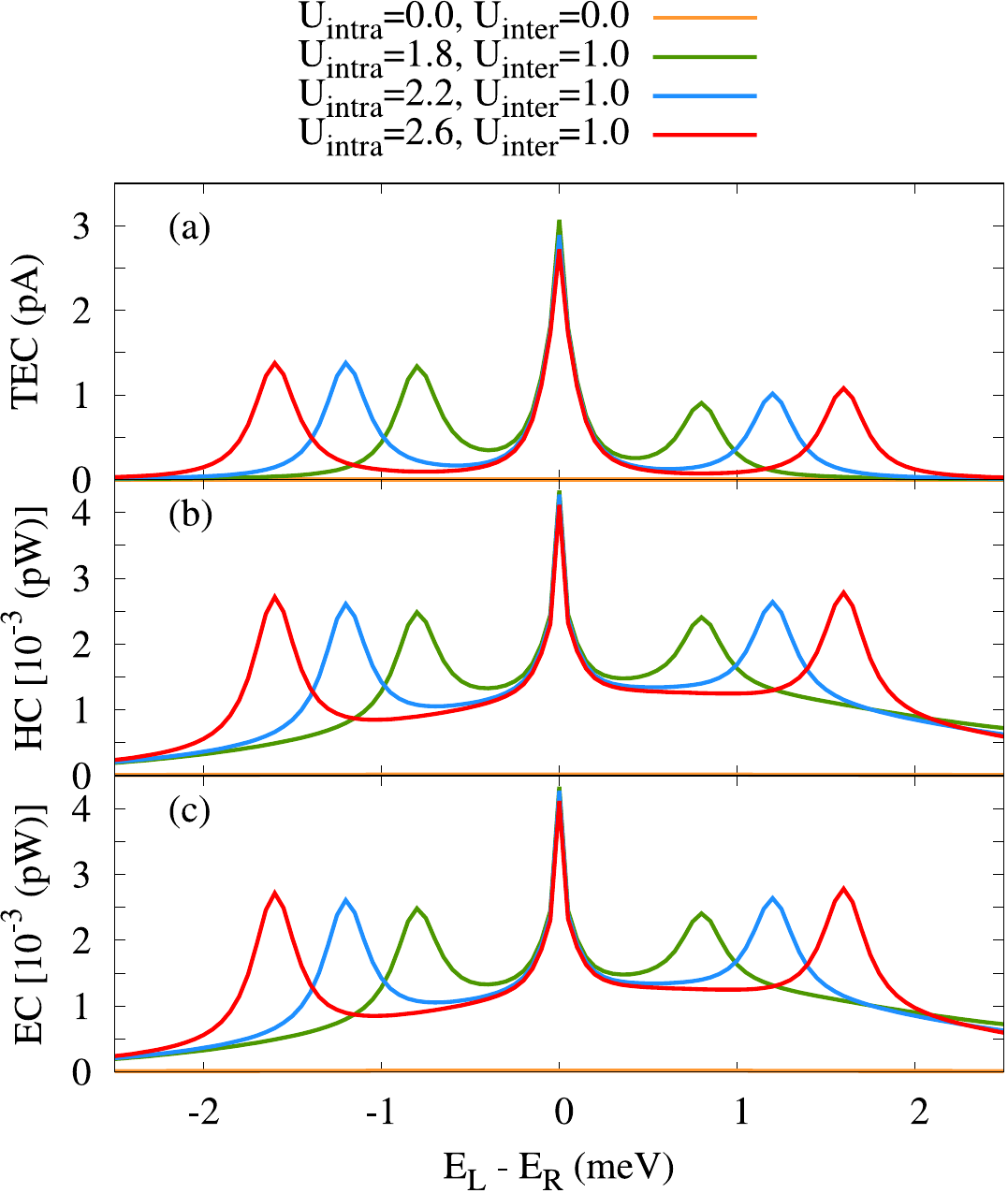}  %MBE_Uintra_0_Uinter_0
	\caption{Thermoelectric current, TEC, (a), Heat current, HC, (b), and Energy current, EC, (c) as a function of $E_{\rm L}\text{-}E_{\rm R}$ for the DQD without Coulomb interactions U$_{\rm inter} = 0.0$~meV, and U$_{\rm intra} = 0.0$~meV (orange), and with intra- and inter-dot Coulomb interactions of U$_{\rm intra} = 1.8$ (green), $2.2$ (blue), $2.6$~meV (red), and 
	the inter-dot Coulomb interaction is fixed at U$_{\rm inter} = 1.0$~meV in all three cases.
	The chemical potentials of the leads are $\mu_L = \mu_R = 0.0$~meV which coincides with the middle vertical golden line when $E_L = E_R$.
    The thermal energy of the leads are assumed to be $k_BT_L = 1.5$~meV and $k_BT_R = 0.5$~meV, and the coupling strength is $\Gamma_{L,R} = 90 \times 10^{-6}$~meV.}
	\label{fig07}
\end{figure}

It is interesting to see the thermoelectric properties in the presence of the intra- and the inter-dot Coulomb interactions. The TEC (a), the HC (b), and the EC (c) are displayed in \fig{fig07} for different strength of the intra-dot Coulomb interactions. 
We compare the DQD without the Coulomb interactions U$_{\rm inter} = 0.0$~meV, and U$_{\rm intra} = 0.0$~meV (orange), and with the intra- and the inter-dot Coulomb interactions U$_{\rm intra} = 1.8$ (green), $2.2$ (blue), $2.6$~meV (red) here, with the inter-dot Coulomb interaction fixed at U$_{\rm inter} = 1.0$~meV in all three cases.
 
Comparing to the thermoelectric properties of the DQD without Coulomb interactions, the TEC, the HC, and the EC are increased over the entire range of $E_{\rm L}\text{-}E_{\rm R}$. The most interesting point here is that the extra side-peaks in the TEC, the HC, and the EC are found in addition to the main peak at $E_{\rm L}\text{-}E_{\rm R}=0.0$. The side peaks are formed at $E_{\rm L}\text{-}E_{\rm R}= \pm \, (U_{\rm intra} \text{-} \, U_{\rm inter}$). 
The current strength for the left and the right side peaks are slightly different arising from different occupation strength at $-(U_{\rm intra} \text{-} \, U_{\rm inter}$) and $+(U_{\rm intra} \text{-} \, U_{\rm inter}$). The origin of the current side peaks comes totally from the contribution of the intra- and the inter-dot 2ESs to the electron transport in the presence of Coulomb interactions.
In fact, the intra- and the inter-dot 2ESs do not only contribute to the current side peak, but they also enhance the main current peak at $E_{\rm L}\text{-}E_{\rm R}=0.0$ as the 2ESs are populated and cross at this point. The side peak in the current is enhanced with increasing strength of the intra-dot Coulomb interaction. The HC and the EC are very similar as the chemical potential of the leads are assumed to be zero, $\mu_{\rm L,R} = 0.0$.

Next we study the effects of cotunneling on the thermoelectric properties of the DQD-leads system. As we mentioned before the 1vN only considers the sequential tunneling between the leads and the DQD. To take into account the cotunneling effect, we consider a second-order von-Neumann master equation, 2vN \cite{PhysRevB.72.195330, PEDERSEN2010595}, which is also implemented in QmeQ.
The approximations for the 2vN method are almost the same as for the 1vN method except terms including up to two electron/hole excitations are taken into account. In addition, the condition for the coupling strength is changed to $\Gamma_{L,R} \lesssim T_{L,R}$ in 2vN method. 
We assume the thermal energy of the leads to be $k_BT_L = 1.5$~meV and $k_BT_R = 0.5$~meV, and the coupling strength is $\Gamma_{L,R} = 50 \times 10^{-5}$~meV. So, our system converges under these assumptions.

\begin{figure}
	\centering
	\includegraphics[width=0.45\textwidth]{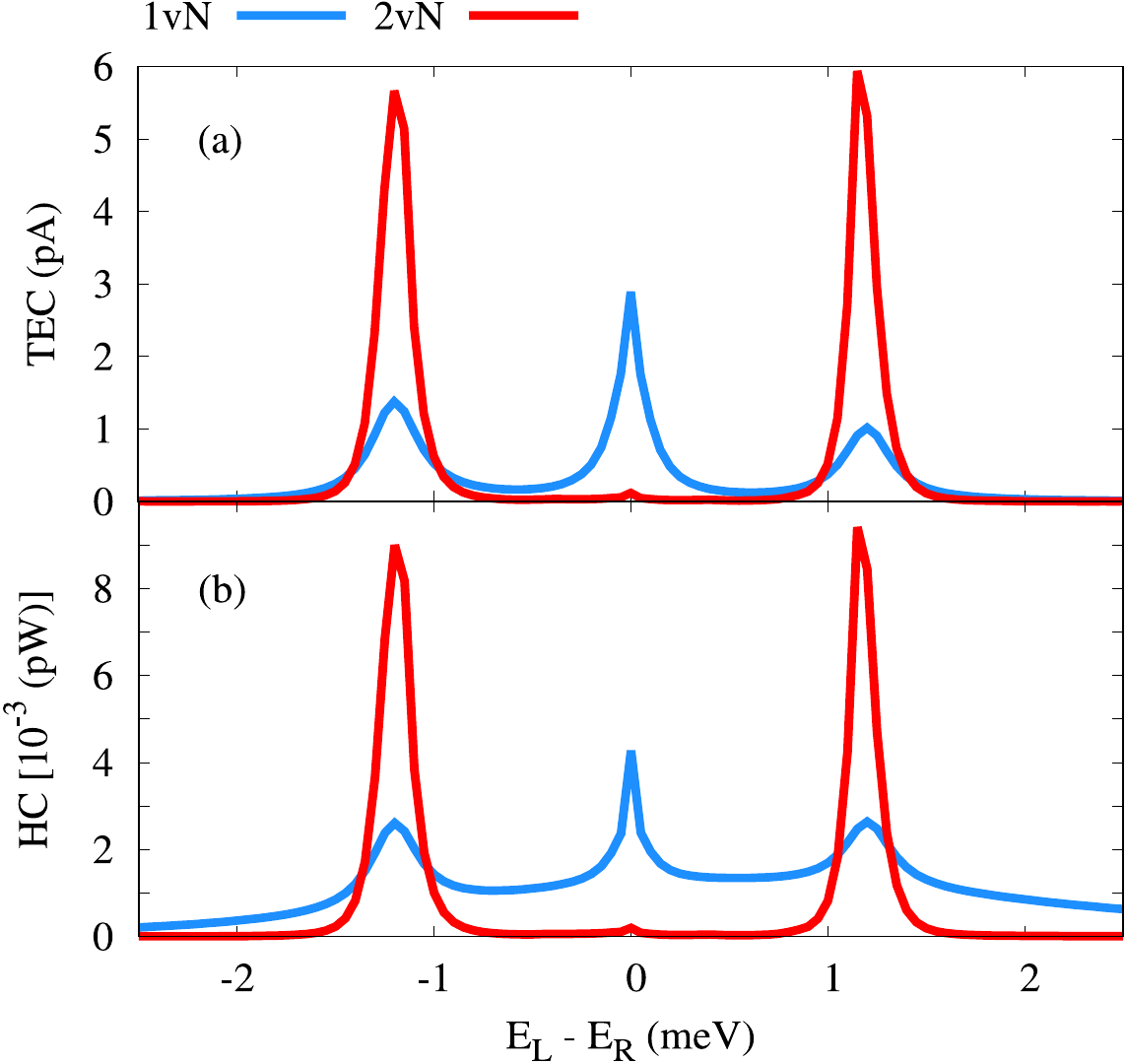} 
	\caption{Thermoelectric current, TEC, (a), and Heat current, HC, (b) as a function of $E_{\rm L}\text{-}E_{\rm R}$ for the DQD where 1vN (blue) and 2vN (red) are considered.
		The intra-dot Coulomb interactions is U$_{\rm intra} = 2.2$~meV, and the inter-dot Coulomb interaction is U$_{\rm inter} =1.0$~meV.
		The chemical potentials of the leads are $\mu_L = \mu_R = 0.0$~meV.
		The thermal energy of the leads are assumed to be $k_BT_L = 1.5$~meV and $k_BT_R = 0.5$~meV, and the coupling strength is $\Gamma_{L,R} = 90 \times 10^{-6}$~meV for 1vN, and $\Gamma_{L,R} = 50 \times 10^{-5}$~meV for 2vN.}
	\label{fig08}
\end{figure}

In \fig{fig08}, we compare the TEC (a), and the HC (b) results obtained via the 1vN (blue), and the 2vN (red), where the Coulomb interactions are U$_{\rm intra} = 2.2$~meV, and U$_{\rm inter} =1.0$~meV. We do not present the the EC here because it is the same as the HC.
It is very interesting to see that the main peak current is extremely suppressed and the side peaks current is strongly enhanced for the 2vN. This is an indication that the crossing of both types of 2ESs at $E_{\rm L}\text{-}E_{\rm R}= \pm \, (U_{\rm intra} \text{-} \, U_{\rm inter}$) can enhance the tunneling of two electrons between the leads and the DQD. As a result, the current of the side peaks is increased.

\begin{figure}
	\centering
	\includegraphics[width=0.45\textwidth]{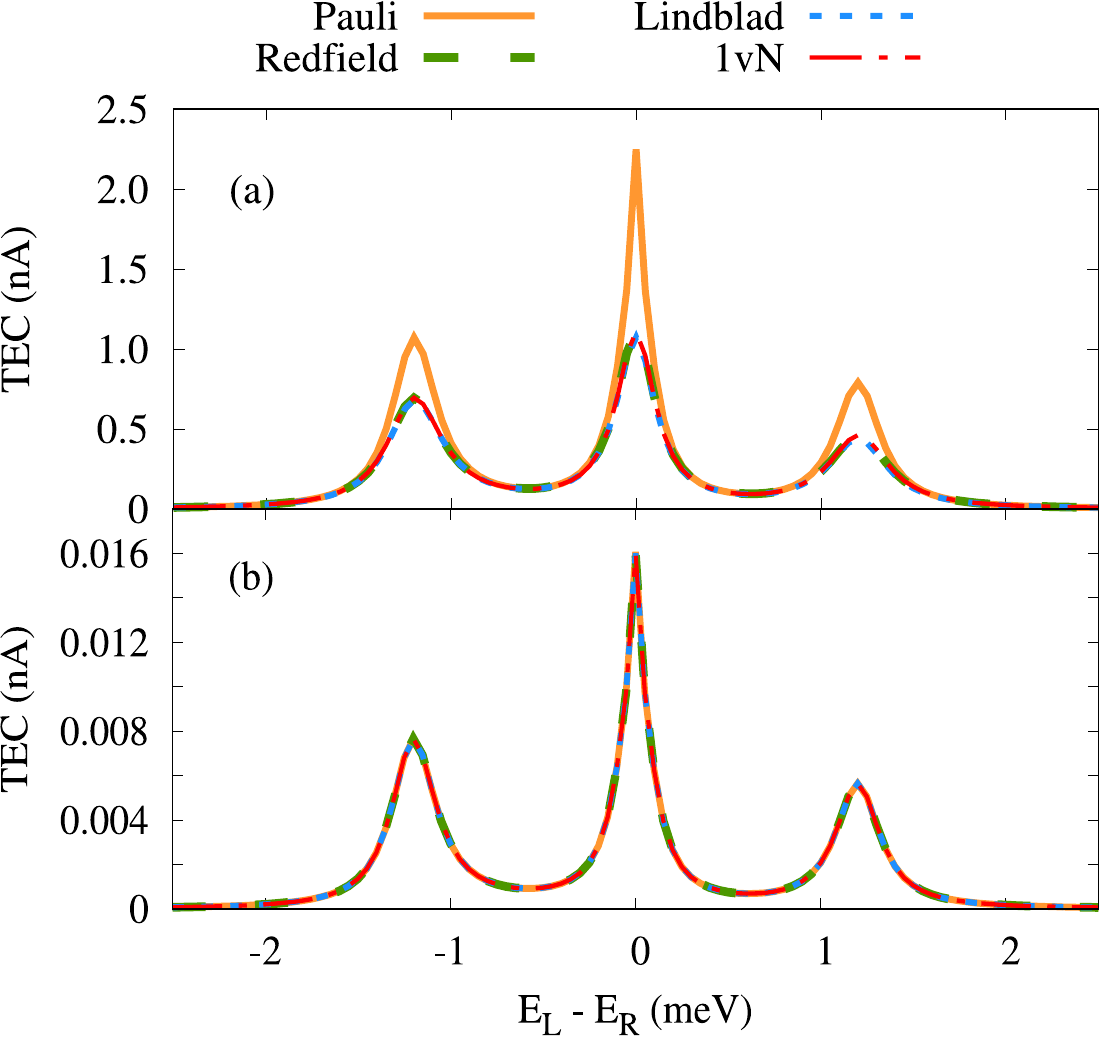} 
	\caption{Thermoelectric current, TEC, as a function of $E_{\rm L} - E_{\rm R}$ for 
	$\Gamma_{\rm L,R} = 0.07 > \Omega = 0.05$ (a), and $\Gamma_{\rm L,R} = 0.0005 < \Omega = 0.05$ (b).
	We assume  U$_{\rm intra} = 2.2$ meV, and U$_{\rm inter} = 1.0$~meV.
    The chemical potentials of the leads are $\mu_L = \mu_R = 0.0$~meV, and
    the thermal energy of the leads are assumed to be $k_BT_L = 1.5$~meV and $k_BT_R = 0.5$~meV}
	\label{fig09}
\end{figure}

Finally, we consider different sequential tunneling approaches of master equations including the Pauli \cite{Grabert1982}, the Redfield \cite{PhysRev.89.728}, and the simple first order Lindblad approach \cite{PEDERSEN2010595, JONSSON201781} in addition to the 1vN method.
In general, the approximations in these sequential approaches of master equations are almost the same except that the sequential tunneling in the presence of coherences is taken
into account for the Redfield, the first order Lindblad approach, and the 1vN approaches, while the coherences in the Pauli formulation are neglected. So, if the coherences are important in a system such as our model, 
we should carefully select and use an appropriate master equation approach. 
The coherences are mainly affected by the inter-dot coupling strength, $\Omega$, and the strength of the coupling between the DQD and the leads, $\Gamma_{\rm L,R}$ \cite{KIRSANSKAS2017317}. 
In \fig{fig09}, the TEC is shown in the case of $\Gamma_{\rm L,R} > \Omega$ (a), and $\Gamma_{\rm L,R} < \Omega$ (b). If $\Gamma_{\rm L,R} > \Omega$, the role of the coherences is relevant, while in the case of $\Gamma_{\rm L,R} < \Omega$, the coherences are irrelevant. We can therefore see that the Pauli master equation gives a wrong TEC in \fig{fig09}(a). In addition, a perfect agreement between the Pauli current and the current obtained via the other methods of master equations is seen when the coherences are irrelevant (see \fig{fig09}(b)).

\section{Conclusion}\label{section_conclusion}

To conclude, the thermoelectric, the heat, and the energy currents through a DQD are studied using five different approaches for a master equation including sequential and cotunneling terms. We take into account the intra- and the inter-dot Coulomb interaction leading to resonance of two different types of two-electron states, 2ES. At the resonance energy of the 2ESs, the intra-dot 2ESs are depopulated and the inter-dot 2ESs are populated. Consequently, the thermal transport is enhanced at these energy resonance leading to the emerging of side peaks in the currents.
If cotunneling processes are considered via a second order master equation (2vN), the current side peaks are enhanced as the crossings of the 2ESs encourage two a electron transport from the leads to the DQD. In addition, the issue of coherences in the sequential tunneling was described and a suitable master equation for thermal transport is proposed when the coherences are relevant in the system. 

\section{Acknowledgment}
This work was financially supported by the University of Sulaimani and 
the Research center of Komar University of Science and Technology. 
The computations were performed on resources provided by the Division of Computational 
Nanoscience at the University of Sulaimani.  
 
%\section{References}

%\bibliographystyle{elsarticle-num} 
%\bibliography{Ref.bib}

\end{document}